\documentclass{eas}
\usepackage{graphicx}
\usepackage{clshan-math,clshan-dm-dd}
\newcommand{\plotsinsert} [5]
{\begin{figure}[#1]
  \begin{center}
   \vspace{-0.25cm}
   {\includegraphics[width=8.2cm]{#2-#3} \\ \vspace{0.5cm}
    \includegraphics[width=8.2cm]{#2-#4} \\
   }
   \vspace{-0.5cm}
  \end{center}
  \caption{#5}
  \label{fig:#2}
 \end{figure}
}
\def \lsim {\:\raisebox{-0.7ex}{$\stackrel{\textstyle<}{\sim}$}\:}

\def \ignore#1 {}
\begin{document}

%
\title{Analyzing Direct Dark Matter Detection Data by the AMIDAS Website}
\runningtitle{Analyzing direct DM detection data by the AMIDAS website}
\author{Chung-Lin Shan}
\address{
 {\it Institute of Physics, Academia Sinica,                 
      No.~128, Sec.~2, Academia Road, Nankang,
      Taipei 11529, Taiwan, R.O.C.}                          \\ 
 {\it E-mail:} {\tt clshan@phys.sinica.edu.tw}               \\
}
%
%
%
\begin{abstract}
 In this talk
 I have presented the data analysis results
 of extracting properties of halo WIMPs:
 the mass and the (ratios between the) spin--independent
 and spin--dependent couplings/cross sections on nucleons
 by the \amidas\ website.
 Although {\em non--standard} astronomical setup
 has been used
 to generate {\em pseudodata} sets for our analyses,
 it has been found that,
 {\em without} prior information/assumption
 about the local density and velocity distribution of halo Dark Matter,
 these WIMP properties have been reconstructed with
 $\sim$ 5\% to $\lsim$ 40\% deviations from the input values.
\end{abstract}
\maketitle
%
%
\section{Introduction}
 In order to extract properties of halo WIMPs
 (Weakly Interacting Massive Particles)
 by using data from direct Dark Matter detection experiments
 as model--independently as possible,
 we have developed a series of data analysis method
 for reconstructing the one--dimensional
 WIMP velocity distribution function
 (Drees \symbol{38} Shan \cite{DMDDf1v})
 as well as determining the WIMP mass
 (Drees \symbol{38} Shan \cite{DMDDmchi}),
 the spin--independent (SI) WIMP coupling on nucleons
 (Shan \cite{DMDDfp2})
 and
 the ratios between different WIMP couplings/cross sections
 (Shan \cite{
             DMDDranap}).
 Moreover,
 in collaboration with
 the DAMNED (DArk Matter Network Exclusion Diagram) Dark Matter online tool
 (\cite{DAMNED}),
 part of the ILIAS Project (\cite{ILIAS}),
 the ``\amidas'' (A Model--Independent Data Analysis System) website
 for online simulation/data analysis
 has also been established
 (\cite{AMIDAS-web};
  Shan \cite{AMIDAS-SUSY09, AMIDAS-f1vFQ}).

 In this article,
 in order to demonstrate the usefulness and powerfulness
 as well as the model--independence of the \amidas\ package
 for direct Dark Matter detection experiments,
 I will analyze {\em blindly} some {\em pseudodata} sets
 generated for different detector materials
 and present the reconstructed WIMP properties.
 This means that
 I will simply upload these data sets onto the \amidas\ website
 and follow the instructions to reconstruct different WIMP properties
 {\em without} using any information about the input setup
 used for generating the pseudodata.
 For cases in which
 some information about WIMPs (e.g., the mass $\mchi$)
 and/or Galactic halo (e.g., the local Dark Matter density $\rho_0$)
 is required,
 I will naively use the commonly used/favorite values
 for the data analyses.

 After that
 I show the blindly reconstructed properties of halo WIMPs
 in Sec.~2,
 in Sec.~3
 I will reveal the input setup
 used for generating the analyzed data
 and compare the reconstructed results to them.
 Finally,
 I conclude
 in Sec.~4.
\section{Reconstructed WIMP properties}
 In this section,
 I present the reconstructed WIMP properties
 analyzed by the \amidas\ website.
 While in each uploaded file
 there are {\em exactly} 50 data sets,
 in each data set
 there are {\em on average} 50 recorded events
 (i.e., 50 measured recoil energies)%
\footnote{
 Note that
 we considered here only data sets with pure WIMP signals,
 possible unrejected background events are neglected.
};
 the exact number of total events is Poisson distributed.
 For simplicity,
 the experimental minimal and maximal cut--off energies
 have been set as 0 and 100 keV
 for all data sets.

 In order to check the effect of
 using a ``wrong'' elastic nuclear form factor,
 two forms have been considered
 for the SI WIMP--nucleus cross section
 in our analyses.
 One is the simple exponential form:
\beq
   F_{\rm ex}^2(Q)
 = e^{-Q / Q_0}
\~.
\label{eqn:FQ_ex}
\eeq
 Here $Q$ is the recoil energy
 transferred from the incident WIMP to the target nucleus,
 $Q_0$ is the nuclear coherence energy given by
\(
   Q_0
 = 1.5 / \mN R_0^2
\),
 where
\(
   R_0
 = \bbig{0.3 + 0.91 \abrac{\mN / {\rm GeV}}^{1/3}}~{\rm fm}
\)
 is the radius of the nucleus
 and $\mN$ is the mass of the target nucleus.
 Meanwhile,
 we used also a more realistic analytic form
 for the elastic nuclear form factor:
\beq
   F_{\rm SI}^2(Q)
 = \bfrac{3 j_1(q R_1)}{q R_1}^2 e^{-(q s)^2}
\~.
\label{eqn:FQ_WS}
\eeq
 Here $j_1(x)$ is a spherical Bessel function,
\(
   q
 = \sqrt{2 m_{\rm N} Q}
\)
 is the transferred 3-momentum,
 for the effective nuclear radius we use
\(
   R_1
 = \sqrt{R_A^2 - 5 s^2}
\)
 with
\(
        R_A
 \simeq 1.2 \~ A^{1/3}~{\rm fm}
\)
 and a nuclear skin thickness
\(
        s
 \simeq 1~{\rm fm}
\).
 For the SD WIMP--nucleus cross section,
 we only used the ``thin--shell'' nuclear form factor:
\beqn
    F_{\rm TS}^2(Q)
 \= \cleft{\renewcommand{\arraystretch}{1.5}
           \begin{array}{l l l}
            j_0^2(q R_1)                      \~, & ~~ &
            {\rm for}~q R_1 \le 2.55~{\rm or}~q R_1 \ge 4.5 \~, \\ 
            {\rm const.} \simeq 0.047         \~, &    &
            {\rm for}~2.5 5 \le q R_1 \le 4.5 \~.
           \end{array}}
\label{eqn:FQ_TS}
\eeqn
\subsection{WIMP mass m$_{\chi}$}
\plotsinsert{t!}{mchi}{ex}{WS}
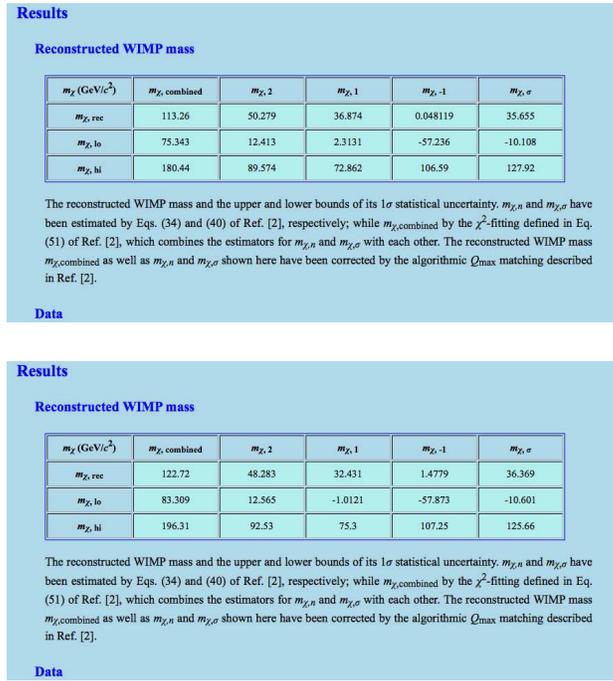
{The WIMP mass $\mchi$
 reconstructed with a target combination of
 $\rmXA{Si}{28}$ + $\rmXA{Ge}{76}$ nuclei.
 Two forms of the elastic nuclear form factor
 given in Eqs.~(\ref{eqn:FQ_ex}) and (\ref{eqn:FQ_WS})
 have been used in the upper and lower frames,
 respectively.
}

 As one of the most important properties of halo WIMPs
 as well as the basic information for reconstructing other quantities
 in our model--independent analysis methods,
 I consider at first
 the determination of the WIMP mass $\mchi$
 by means of the method introduced
 in Drees \symbol{38} Shan (\cite{DMDDmchi}).

 In Figs.~\ref{fig:mchi}
 I show the reconstructed WIMP masses and
 the upper and lower bounds of their 1$\sigma$ statistical uncertainties.
 The usual target combination of
 \mbox{$\rmXA{Si}{28}$ + $\rmXA{Ge}{76}$} nuclei
 has been used for this reconstruction,
 whereas
 two forms of the elastic nuclear form factor
 given in Eqs.~(\ref{eqn:FQ_ex}) and (\ref{eqn:FQ_WS})
 have been used for determining $\mchi$
 in the upper and lower frames,
 respectively.
 While $m_{\chi, n}$ with $n = -1$, 1, 2 and $m_{\chi, \sigma}$
 have been estimated by Eqs.~(34) and (40) of Drees \symbol{38} Shan (\cite{DMDDmchi}),
 respectively,
 $m_{\chi, {\rm combined}}$ has been estimated
 by the $\chi^2$--fitting
 defined in Eq.~(51) of Drees \symbol{38} Shan (\cite{DMDDmchi}),
 which combines the estimators for
 $m_{\chi, n}$ and $m_{\chi, \sigma}$ with each other.
 The reconstructed WIMP mass $m_{\chi, {\rm combined}}$
 as well as $m_{\chi, n}$ and $m_{\chi, \sigma}$
 shown here have been corrected
 by the iterative $\Qmax$--matching procedure
 described in Drees \symbol{38} Shan (\cite{DMDDmchi}).

 It can be found here that,
 although all single estimators
 ($m_{\chi, n}$ with $n = -1$, 1, 2 and $m_{\chi, \sigma}$)
 give generally a (relatively lighter) WIMP mass of
 \mbox{$\sim 50$ GeV} or even lighter
 and a 1$\sigma$ upper bound of $\sim 130$ GeV,
 the {\em mean} values of the combined
 (in principle, more reliable) results
 (the second column in two tables)
 of the reconstructed WIMP mass give $\mchi \sim 120$ GeV
 with a {\em rough} 1$\sigma$ upper (lower) bound of $\sim 190$ (80) GeV,
 or, equivalently,
\beq
        \mchi
 \simeq 120_{-40}^{+70}~{\rm GeV}
\~.
\label{eqn:mchi_rec}
\eeq
 Moreover,
 the combined results
 with two different form factors
 show not only a large overlap between $\sim 85$ GeV and $\sim 180$ GeV,
 but also a good coincidence:
 comparing to the $\sim_{-40}^{+70}$ GeV
 1$\sigma$ statistical uncertainty
 and the $\sim_{-35}^{+60}$ GeV overlap,
 the difference between two median values is $\lsim~10$ GeV!
 This indicates that,
 for the first approximation of
 giving/constraining the most plausible range of the WIMP mass,
 the uncertainty on the nuclear form factor
 could be safely neglected.

\subsection{Spin--independent WIMP--nucleon coupling $|$f$_{\sf p}|^2$}

 Following the WIMP mass determination,
 I consider now the reconstruction of
 the SI WIMP coupling on nucleons $|f_{\rm p}|^2$
 (Shan \cite{DMDDfp2})
 with a $\rmXA{Ge}{76}$ target%
\footnote{
 Remind that
 the theoretical prediction
 by most supersymmetric models that
 the SI scaler WIMP couplings
 on protons and on neutrons are (approximately) equal:
 $f_{\rm p} \simeq f_{\rm n}$
 has been adopted in the \amidas\ package.
}.
\plotsinsert{t!}{fp2}{03-ex}{input-WS}
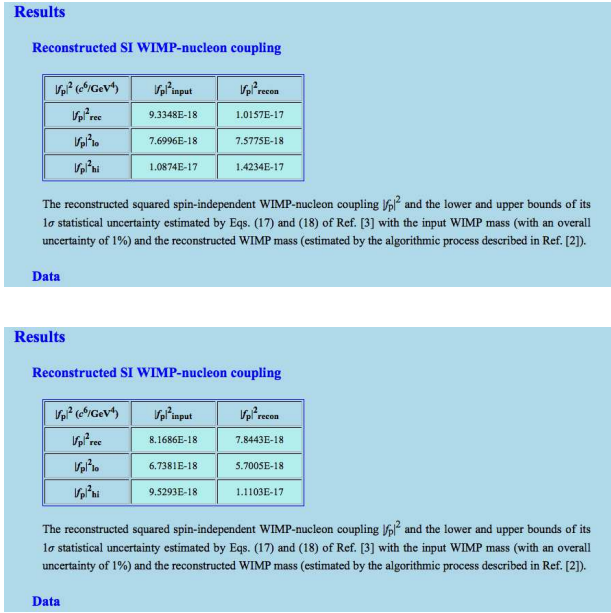
{The {\em squared} SI WIMP--nucleon coupling $|f_{\rm p}|^2$
 reconstructed with a $\rmXA{Ge}{76}$ target.
 The commonly used value of
 the local Dark Matter density
 $\rho_0 = 0.3~{\rm GeV/cm^3}$
 and a larger value of $\rho_0 = 0.4~{\rm GeV/cm^3}$
 as well as
 the elastic nuclear form factors
 given in Eqs.~(\ref{eqn:FQ_ex}) and (\ref{eqn:FQ_WS})
 have been used for estimating $|f_{\rm p}|^2$
 in the upper and lower frames,
 respectively.
}

 In Figs.~\ref{fig:fp2}
 I show the reconstructed squared SI WIMP-nucleon couplings
 and the lower and upper bounds of
 their 1$\sigma$ statistical uncertainties
 estimated by Eqs.~(17) and (18) of Shan (\cite{DMDDfp2})
 with an {\em assumed} (100$\pm$10 GeV, labeled with the subscript ``input'')
 and the reconstructed (from Sec.~2.1, labeled with ``recon'') WIMP masses.
 The commonly used value of
 the local Dark Matter density
 $\rho_0 = 0.3~{\rm GeV/cm^3}$
 and a larger value of $\rho_0 = 0.4~{\rm GeV/cm^3}$
 (Catena \symbol{38} Ullio \cite{Catena09};
  Salucci {\it et al.}~\cite{Salucci10};
  Pato {\it et al.}~\cite{Pato10})
 as well as
 the elastic nuclear form factors
 given in Eqs.~(\ref{eqn:FQ_ex}) and (\ref{eqn:FQ_WS})
 have been used for estimating $|f_{\rm p}|^2$
 in the upper and lower frames,
 respectively.

 Among these results,
 the {\em mean} value and the {\em overlap} of two most plausible results
 (estimated by using the reconstructed WIMP mass)
 give {\em roughly} (and somehow {\em naively}) a 1$\sigma$ range of
\beq
        |f_{\rm p}|^2
 \simeq 9.00_{-1.44}^{+2.10} \times 10^{-18}~{\rm GeV}^{-4}
\~,
\label{eqn:fp2_rec}
\eeq
 or, equivalently,
\beq
        |f_{\rm p}|
 \simeq 3.00_{-0.24}^{+0.35} \times 10^{-9}~{\rm GeV}^{-2}
\~.
\label{eqn:fp_rec}
\eeq
 Since the reconstructed WIMP mass given in Sec.~2.1 is
 $\mchi \sim 120$ GeV,
 one can simply use the proton mass $m_{\rm p}$
 to approximate the WIMP--proton reduced mass $\mrp$
 and give a reconstructed SI WIMP--nucleon cross section as%
\footnote{
 Note that,
 since the expression for estimating $|f_{\rm p}|^2$
 (Eq.~(17) of Shan (\cite{DMDDfp2}))
 is a function of the (reconstructed) WIMP mass,
 for light WIMP mass,
 one has to use
\beq
   \sigmapSI
 = \frac{1}{\rho_0} \!\!
   \bbrac{\frac{1}{\sqrt{2}} \afrac{1}{\calE A^2 \sqrt{\mN}}} \!\!
   \bbrac{\frac{2 \Qmin^{1/2} r(\Qmin)}{\FQmin} + I_0} \!
   \abrac{\mchi + \mN} \!
   \afrac{\mchi m_{\rm p}}{\mchi + m_{\rm p}}^2 \!\!
,
\label{eqn:sigmapSI_mchi}
\eeq
 where $A$ is the atomic mass number of the target nucleus,
 $\calE$ is the experimental exposure.
 Then one has (cf.~Eq.~(18) of Shan (\cite{DMDDfp2}))
\beqn
   \sigma\abrac{\sigmapSI}
 \&=\&
   \sigmapSI
   \cleft{   \frac{\sigma^2(\mchi)}{(\mchi + \mN)^2} \bbig{1 + \Delta(\mchi)}^2
           + \calN_{\rm m}^2 \sigma^2(1 / \calN_{\rm m}) }
   \non\\
 \&~\& ~~~~~~~~~~~~~~~~~~~~~~~~ 
   \cright{+ \frac{2 \calN_{\rm m} \~ {\rm cov}(\mchi, 1 / \calN_{\rm m})}
                  {(\mchi + \mN)} \bbig{1 + \Delta(\mchi)} }^{1/2}
\~.
\label{eqn:sigma_sigmapSI_mchi}
\eeqn
 Here I have used (Drees \symbol{38} Shan \cite{DMDDf1v})
\beq
   \calN_{\rm m}
 = \bbrac{\frac{2 \Qmin^{1/2} r(\Qmin)}{\FQmin} + I_0}^{-1}
\~,
\label{eqn:calNm}
\eeq
 and defined
\beq
   \Delta(\mchi)
 = 2 \afrac{m_{\rm p}}{\mchi} \afrac{\mchi + \mN}{\mchi + m_{\rm p}}
\~.
\label{eqn:Delta_mchi}
\eeq
 Definitions and estimations of $r(\Qmin)$ and $I_n$
 can be found in e.g.,~Drees \symbol{38} Shan (\cite{DMDDf1v, DMDDmchi}).
}
\beq
         \sigmapSI
 =       \afrac{4}{\pi} \mrp^2 |f_{\rm p}|^2
 \approx \afrac{4}{\pi} m_{\rm p}^2 \~ |f_{\rm p}|^2
 \simeq  4.31_{-0.69}^{+1.01} \times 10^{-9}~{\rm pb}
\~.
\label{eqn:sigmapSI_rec}
\eeq
\subsection{Ratio of two spin--dependent WIMP--nucleon couplings a$_{\sf n}$/a$_{\sf p}$}
\plotsinsert{t!}{ranap}{ex}{WS}
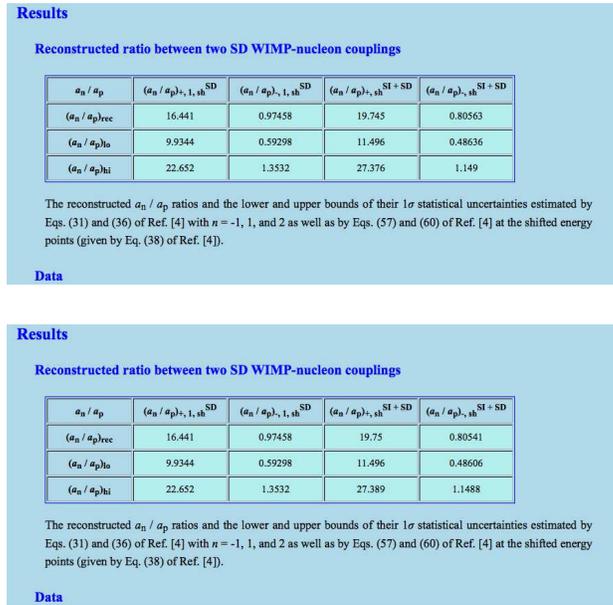
{The reconstructed ratio between two SD WIMP--nucleon couplings,
 $\armn / \armp$.
 As usual,
 the elastic nuclear form factors
 given in Eqs.~(\ref{eqn:FQ_ex}) and (\ref{eqn:FQ_WS})
 have been used for determining $\armn / \armp$
 in the upper and lower frames,
 respectively.
}

 In Figs.~\ref{fig:ranap}
 I show the reconstructed $\armn / \armp$ ratios
 and the lower and upper bounds of
 their 1$\sigma$ statistical uncertainties
 estimated by Eqs.~(2.7) and (2.12) of Shan (\cite{DMDDranap}) with $n = 1$
 as well as by Eqs.~(3.16) and (3.20) of Shan (\cite{DMDDranap})
 at the shifted energy points
 (Drees \symbol{38} Shan \cite{DMDDf1v}; Shan \cite{DMDDranap}).
 A combination of $\rmXA{F}{19}$ + $\rmXA{I}{127}$ targets
 has been used
 for the reconstruction of $\armn / \armp$
 under the assumption that
 the SD WIMP--nucleus interaction dominates over the SI one
 (labeled with the superscript ``SD''),
 whereas
 a third target of $\rmXA{Si}{28}$ has been combined
 with $\rmXA{F}{19}$ and $\rmXA{I}{127}$
 for the case of the general combination of both SI and SD WIMP interactions
 (labeled with the superscript ``SI + SD'').

 It can be found that,
 firstly,
 the ``$+$ (plus)'' solutions of the $\armn / \armp$ ratios given here
 are obviously too large to be the reasonable choice for $\armn / \armp$
 and the ``$-$ (minus)'' solutions should be the correct ones%
\footnote{
 Remind that,
 as discussed in Shan (\cite{
                             DMDDranap}),
 the correct choice from the ``$+$'' and ``$-$'' solutions
 can be decided directly by the values of
 the group spins of protons and neutrons of the used target nuclei,
 $\expv{S_{\rm (p, n)}}$.
}.
 Secondly,
 although the reconstructed result
 under the assumption of the SD dominant WIMP interaction
 is in general {\em larger} than
 the (in principle more plausible) result
 obtained without such a prior assumption%
\footnote{
 See also discussions in Sec.~2.4.
},
 one could still use
 the {\em mean} value and the {\em overlap} of these two results
 to {\em roughly} (and somehow {\em naively}) give a 1$\sigma$ range of
\beq
        \frac{\armn}{\armp}
 \simeq 0.89_{-0.30}^{+0.26}
\~.
\label{eqn:ranap_rec}
\eeq
\subsection{Ratios of the SD and SI WIMP--nucleon couplings
            $\sigma_{\chi ({\sf p, n})}^{\sf SD} / \sigma_{\chi {\sf p}}^{\sf SI}$}
\plotsinsert{t!}{rsigmaSDSI}{ex}{WS}
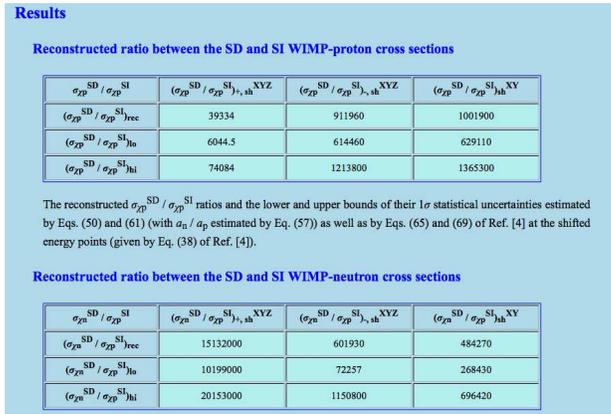
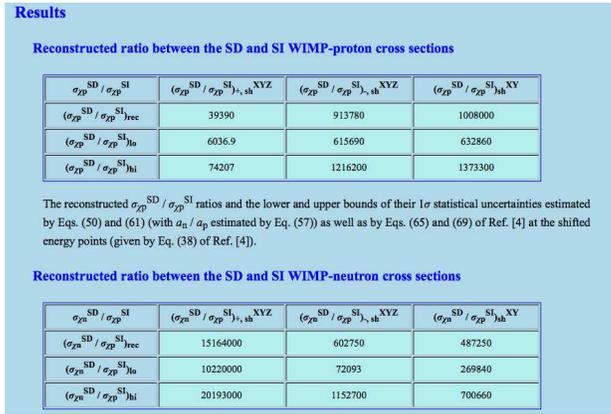
{The reconstructed ratios between the SD and SI WIMP--nucleon couplings,
 $\sigma_{\chi ({\rm p, n})}^{\rm SD} / \sigmapSI$.
 As usual,
 the elastic nuclear form factors
 given in Eqs.~(\ref{eqn:FQ_ex}) and (\ref{eqn:FQ_WS})
 have been used for determining
 $\sigma_{\chi ({\rm p, n})}^{\rm SD} / \sigmapSI$
 in the upper and lower frames,
 respectively.
}

 In Figs.~\ref{fig:rsigmaSDSI}
 I show the reconstructed
 $\sigma_{\chi ({\rm p, n})}^{\rm SD} / \sigmapSI$ ratios
 and the lower and upper bounds of
 their 1$\sigma$ statistical uncertainties
 estimated by Eqs.~(3.9), (3.10) and (3.21) of Shan (\cite{DMDDranap})
 (with $\armn / \armp$ estimated by Eq. (3.16) of Shan (\cite{DMDDranap}))
 as well as by Eqs.~(3.25) and (3.29) of Shan (\cite{DMDDranap})
 at the shifted energy points.

 By using the data sets of
 $\rmXA{F}{19}$, $\rmXA{I}{127}$ and $\rmXA{Si}{28}$ targets
 (labeled with the superscript ``XYZ'')
 or combining that of $\rmXA{Na}{23}$ or $\rmXA{Xe}{131}$
 with the (common) data set of $\rmXA{Ge}{76}$
 (labeled with the superscript ``XY''),
 one can use
 the {\em mean} value and the {\em overlap} of these two results
 to {\em roughly} (and somehow {\em naively}) give a 1$\sigma$ range of
\beq
        \frac{\sigmapSD}{\sigmapSI}
 \simeq 9.61_{-3.28}^{+2.55} \times 10^5
\~,
 ~~~~~~~~~~~~ 
        \frac{\sigmanSD}{\sigmapSI}
 \simeq 5.45_{-2.75}^{+1.56} \times 10^5
\~.
\label{eqn:rsigmaSDSI_rec}
\eeq
 Then,
 firstly,
 from these results one can further obtain that%
\footnote{
 Here I have used
\beqn
   \sigma\abrac{\vfrac{\armn}{\armp}}
 \&=\&
   \frac{1}{2} \vfrac{\armn}{\armp}
   \bleft{   \sigma^2\aBig{\sigmanSD / \sigmapSI}   \left/
                     \aBig{\sigmanSD / \sigmapSI}^2 \right. }
   \non\\
 \&~\& ~~~~~~~~~~~~~~~~~~~~~~~~ 
   \bright{+ \sigma^2\aBig{\sigmapSD / \sigmapSI}   \left/
                     \aBig{\sigmapSD / \sigmapSI}^2 \right. }^{1/2}
\~,
\label{eqn:sigma_ranap_2}
\eeqn
 and neglected the correlation term in the bracket:
\beq
 -2 \~
  {\rm cov}\aBig{\sigmanSD / \sigmapSI,       \sigmapSD / \sigmapSI} \left/
           \aBig{\sigmanSD / \sigmapSI} \aBig{\sigmapSD / \sigmapSI} \right.
\~,
\label{eqn:sigma_ranap_2_cov}
\eeq
 since the 1$\sigma$ uncertainties
 given in Eq.~(\ref{eqn:rsigmaSDSI_rec})
 are not the exact but only rough estimates
 from the overlaps of two results given in Figs.~\ref{fig:rsigmaSDSI}.
}$^{,~}$%
\footnote{
 Remind that
 the results given in the second and third columns
 of the tables in Figs.~\ref{fig:rsigmaSDSI}
 are reconstructed with the $\armn / \armp$ ratio
 given in the last columns of the tables in Figs.~\ref{fig:ranap}.
}
\beq
        \vfrac{\armn}{\armp}
 \simeq 0.75_{-0.23}^{+0.15}
\~.
\label{eqn:ranap_rec_2}
\eeq
 Secondly,
 combining the results
 in Eq.~(\ref{eqn:rsigmaSDSI_rec})
 with $\sigmapSI$ given in Eq.~(\ref{eqn:sigmapSI_rec}),
 one can also obtain that%
\footnote{
 Here I have used
\beq
   \sigma\abrac{\sigma_{\chi {\rm (p, n)}}^{\rm SD}}
 = \bbrac{  \abrac{\sigmapSI}^2
            \sigma^2\abrac{\sigma_{\chi {\rm (p, n)}}^{\rm SD} / \sigmapSI}
          + \abrac{\sigma_{\chi {\rm (p, n)}}^{\rm SD} / \sigmapSI}^2
            \sigma^2\abrac{\sigmapSI}}^{1/2}
\~,
\label{eqn:sigma_sigmapSD}
\eeq
 and neglected the correlation term in the bracket:
\beq
 2 \~
 {\rm cov}\abrac{\sigmapSI,        \sigma_{\chi {\rm (p, n)}}^{\rm SD} / \sigmapSI} \left/
          \abrac{\sigmapSI} \abrac{\sigma_{\chi {\rm (p, n)}}^{\rm SD} / \sigmapSI} \right.
\label{eqn:sigma_sigmapSD_cov}
\eeq
 by assuming that
 two independent data sets with the $\rmXA{Ge}{76}$ target
 and other two independent data sets with the $\rmXA{Si}{28}$ target
 have been used for determining $\sigmapSI$
 and $\sigma_{\chi {\rm (p, n)}}^{\rm SD} / \sigmapSI$.
}
\beq
        \sigmapSD
 \simeq 4.14_{-1.56}^{+1.47} \times 10^{-3}~{\rm pb}
\~,
 ~~~~~~~~~~~~ 
        \sigmanSD
 \simeq 2.35_{-1.31}^{+0.87} \times 10^{-3}~{\rm pb}
\~.
\label{eqn:rsigmaSD_rec}
\eeq
 These results give in turn that%
\footnote{
 Since
\beq
   \sigma_{\chi {\rm (p, n)}}^{\rm SD}
 = \afrac{24}{\pi} G_F^2 \~ m_{\rm r, (p, n)}^2 |a_{\rm (p, n)}|^2
\label{eqn:sigmap/nSD}
\eeq
 and $\mchi \sim 120$ GeV,
 one has
\beq
   |a_{\rm (p, n)}|
 = \sfrac{\pi}{24}
   \frac{\sqrt{\sigma_{\chi {\rm (p, n)}}^{\rm SD}}}{G_F \~ m_{\rm r, (p, n)}}
 \approx \sfrac{\pi}{24}
   \frac{\sqrt{\sigma_{\chi {\rm (p, n)}}^{\rm SD}}}{G_F \~ m_{\rm (p, n)}}
\~,
\label{eqn:an/p}
\eeq
 and
\beq
   \sigma\abrac{|a_{\rm (p, n)}|}
 = \sfrac{\pi}{96}
   \frac{\sigma\abig{\sigma_{\chi {\rm (p, n)}}^{\rm SD}}}
        {G_F \~ m_{\rm r, (p, n)} \sqrt{\sigma_{\chi {\rm (p, n)}}^{\rm SD}}}
 \approx
   \sfrac{\pi}{96}
   \frac{\sigma\abig{\sigma_{\chi {\rm (p, n)}}^{\rm SD}}}
        {G_F \~ m_{\rm (p, n)}    \sqrt{\sigma_{\chi {\rm (p, n)}}^{\rm SD}}}
\~.
\label{eqn:sigma_an/p}
\eeq
}
\beq
        |\armp|
 \simeq 0.108_{-0.020}^{+0.019}
\~,
 ~~~~~~~~~~~~ 
        |\armn|
 \simeq 0.081_{-0.023}^{+0.015}
\~.
\label{eqn:a_rec}
\eeq
 On the other hand,
 one can also use the reconstructed $\armn / \armp$ ratio
 given in Eq.~(\ref{eqn:ranap_rec})
 and {\em one} of the two results
 given in Eq.~(\ref{eqn:rsigmaSD_rec})
 to obtain that%
\footnote{
 Here I have used
\beq
   \sigma\abrac{\sigma_{\chi {\rm (p, n)}}^{\rm SD}}
 = \sigma_{\chi {\rm (p, n)}}^{\rm SD}
   \bbrac{  \sigma^2\abrac{ \sigma_{\chi {\rm (n, p)}}^{\rm SD} }   \left/
                    \abrac{ \sigma_{\chi {\rm (n, p)}}^{\rm SD} }^2 \right.
          + 4 \sigma^2\abig{ \armn / \armp }   \left/
                      \abig{ \armn / \armp }^2 \right. }^{1/2}
\~,
\label{eqn:sigma_sigmapSD_2}
\eeq
 and neglected the correlation term in the bracket:
\beq
 \mp 4 \~
 {\rm cov}\aBig{\sigma_{\chi {\rm (n, p)}}^{\rm SD},       \armn / \armp} \left/
          \aBig{\sigma_{\chi {\rm (n, p)}}^{\rm SD}} \abig{\armn / \armp} \right.
\~,
\label{eqn:sigma_sigmapSD_2_cov}
\eeq
 since the 1$\sigma$ uncertainties
 given in Eq.~(\ref{eqn:ranap_rec}) as well as
 in Eq.~(\ref{eqn:rsigmaSD_rec})
 are not the exact but only rough estimates
 from the overlaps of the results
 given in Figs.~\ref{fig:fp2}, \ref{fig:ranap} and \ref{fig:rsigmaSDSI}.
 The ``$-$ ($+$)'' sign in Eq.~(\ref{eqn:sigma_sigmapSD_2_cov})
 is for the case with protons (neutrons).
}
\beq
        \sigmapSD
 \simeq 2.97_{-2.60}^{+2.05} \times 10^{-3}~{\rm pb}
\~,
 ~~~~~~~~~~~~ 
        \sigmanSD
 \simeq 3.28_{-2.53}^{+2.24} \times 10^{-3}~{\rm pb}
\~.
\label{eqn:rsigmaSD_rec_2}
\eeq
 These results can also give that
\beq
        |\armp|
 \simeq 0.091_{-0.040}^{+0.032}
\~,
 ~~~~~~~~~~~~ 
        |\armn|
 \simeq 0.096_{-0.037}^{+0.033}
\~.
\label{eqn:a_rec_2}
\eeq

 It can be found that,
 not surprisingly,
 the statistical uncertainties
 on the reconstructed $\sigma_{\chi ({\rm p, n})}^{\rm SD}$
 given in Eq.~(\ref{eqn:rsigmaSD_rec_2})
 are $\sim$ 2 or 3 times larger than those
 given in Eq.~(\ref{eqn:rsigmaSD_rec}):
 Since $\sigma_{\chi ({\rm p, n})}^{\rm SD} / \sigmapSI$
 reconstructed with the F + I + Si combination
 involve already the reconstructed $\armn / \armp$ ratio
 given in Eq.~(\ref{eqn:ranap_rec}),
 the uncertainties on $\sigma_{\chi ({\rm p, n})}^{\rm SD}$
 given in Eq.~(\ref{eqn:rsigmaSD_rec_2})
 are thus {\em overestimated}.
 Secondly,
 although the reconstructed $\sigmapSD$ and $\sigmanSD$
 given in Eqs.~(\ref{eqn:rsigmaSD_rec}) and (\ref{eqn:rsigmaSD_rec_2})
 have overlaps,
 these results seem {\em not} to match to each other very well;
 $\sigmanSD$ given in Eq.~(\ref{eqn:rsigmaSD_rec_2})
 is even larger than $\sigmapSD$ there
 although the $\armn / \armp$ ratios
 given in Eqs.~(\ref{eqn:ranap_rec}) and (\ref{eqn:ranap_rec_2})
 are $<$ 1.
 One possible explanation is that
 the $\armn / \armp$ ratio given in Eq.~(\ref{eqn:ranap_rec})
 would be {\em overestimated}.
 This can be seen
 by comparing the $\armn / \armp$ ratio given in Eq.~(\ref{eqn:ranap_rec})
 to that given in Eq.~(\ref{eqn:ranap_rec_2})
 estimated (somehow {\em independently}) by the results
 given in Eq.~(\ref{eqn:rsigmaSDSI_rec}).

 Nevertheless,
 the analyses given here show that,
 firstly,
 once one can estimate
 the SI WIMP--nucleon coupling/cross section,
 $|f_{\rm p}|$ or $\sigmapSI$,
 and (one of) the ratios between
 the SD and SI WIMP--nucleon cross sections,
 and/or the ratio between two SD WIMP--nucleon couplings,
 the other couplings/cross sections
 could in principle be estimated.
 Secondly,
 although the method
 under the assumption of the SD dominant WIMP interaction
 would overestimate
 (or underestimate,
  depending on the combination of the used targets (Shan \cite{DMDDranap}))
 the $\armn / \armp$ ratio,
 the reconstructed result(s)
 could still be useful for at least determining
 the correct sign of $\armn / \armp$.
 Moreover,
 the WIMP couplings/cross sections estimated in different way
 would be self--cross--checks to each other and
 the (in)compatibility between the reconstructed results
 would also help us to check
 the usefulness of the analyzed data sets
 offered from different experiments with different detector materials.

\section{Input setup for generating pseudodata}

 In Table \ref{tab:setup}
 I give finally the input setup
 for generating the pseudodata sets used in the analyses
 demonstrated in the previous section.
 For comparison,
 the reconstructed results shown in the previous section
 are also summarized here.

 It can be found that,
 firstly,
 not only the WIMP mass given in Eq.~(\ref{eqn:mchi_rec})
 and the result reconstructed with the input nuclear form factor
 (lower frame of Figs.~\ref{fig:mchi}),
 but even the mass reconstructed with the ``wrong'' form factor
 (upper frame)
 can match the input WIMP mass very well:
 the deviations between the input and the reconstructed values
 are only $\sim$ 13\% (with the wrong nuclear form factor)
 or even only $\sim$ 6\% (with the input one).
 As discussed earlier,
 this indicates that,
 for the first approximation of
 giving/constraining the most plausible range of the WIMP mass,
 the uncertainty on the nuclear form factor
 could be safely neglected.

\begin{table}[t]
\begin{center}
\renewcommand{\arraystretch}{0.23}
\begin{tabular}{|| c | c | c | l ||}
\hline
\hline
 & & & \\
 \makebox[1.25cm][c]{Property}              &
 \makebox[3.75cm][c]{Reconstructed   value} &
 \makebox[3.75cm][c]{Input/Estimated value} &
 \makebox[1.25cm][c]{Remarks}               \\
 & & & \\
\hline
\hline
 & & & \\
 $\mchi$                   &
 $120_{-40}^{+70}$ GeV     &
  130              GeV     & \\
 & & & \\
\hline
\hline
 & & & \\
 $\sigmapSI$               &
 $4.31_{-0.69}^{+1.01}     \times 10^{-9}$ pb &
 $4                        \times 10^{-9}$ pb &
 $f_{\rm n} = f_{\rm p}$   \\
 & & & \\
\hline
 & & & \\
 $|f_{\rm p}|^2$           &
 $9.00_{-1.44}^{+2.10}     \times 10^{-18}~{\rm GeV}^{-4}$ &
 $9.305                    \times 10^{-18}~{\rm GeV^{-4}}$ &
 $\dagger$                 \\
 & & & \\
\hline
 & & & \\
 $|f_{\rm p}|$             &
 $3.00_{-0.24}^{+0.35}     \times 10^{-9 }~{\rm GeV}^{-2}$ &
 $3.050                    \times 10^{-9 }~{\rm GeV}^{-2}$ &
 $\dagger$                 \\
 & & & \\
\hline
\hline
 & & & \\
 $\armp$                   &
 $0.108_{-0.020}^{+0.019}$ &
  0.1                      & \\
 & & & \\

 $\armn$                   &
 $0.081_{-0.023}^{+0.015}$ &
  0.07                     & \\
 & & & \\
\hline
 & & & \\
 $\armn / \armp$           &
 $0.89_{-0.30}^{+0.26}$,
 $0.75_{-0.23}^{+0.15}$    &
  0.7                      & \\
 & & & \\
\hline
 & & & \\
 $\sigmapSD$               &
 $4.14_{-1.56}^{+1.47}     \times 10^{-3}$ pb &
 $3.51                     \times 10^{-3}$ pb &
 $\dagger$                 \\
 & & & \\
 & & & \\
 $\sigmanSD$               &
 $2.35_{-1.31}^{+0.87}     \times 10^{-3}$ pb &
 $1.72                     \times 10^{-3}$ pb &
 $\dagger$                 \\
 & & & \\
\hline
 & & & \\
 $\sigmapSD / \sigmapSI$   &
 $9.61_{-3.28}^{+2.55}     \times 10^5$ &
 $8.77                     \times 10^5$ &
 $\dagger$                 \\
 & & & \\
 & & & \\
 $\sigmanSD / \sigmapSI$   &
 $5.45_{-2.75}^{+1.56}     \times 10^5$ &
 $4.30                     \times 10^5$ &
 $\dagger$                 \\
 & & & \\
\hline
\hline
 & & & \\
 $\FSIQ$                   & & $\FSIQ$ in Eq.~(\ref{eqn:FQ_WS})           & \\
 & & & \\
 $\FSDQ$                   & & $F_{\rm TS}^2(Q)$ in Eq.~(\ref{eqn:FQ_TS}) & \\
 & & & \\
\hline
\hline
 & & & \\
 $\rho_0$            & & $0.4~{\rm GeV/cm^3}$ & \\
 & & & \\
\hline
 & & & \\
 $t_{\rm p}$         & & 140 d                & \\
 & & & \\
 $t_{\rm expt}$      & & 300 d                & \\
 & & & \\
\hline
 & & & \\
 $v_0$               & & $230~{\rm km/s}$     & \\
 & & & \\
 $\vmax$             & & $600~{\rm km/s}$     & \\
 & & & \\
 $\ve(t_{\rm expt})$ & & $226.6~{\rm km/s}$   & \\
 & & & \\
\hline
\hline
\end{tabular}
\caption{
 The input setup for generating the pseudodata sets
 used in the analyses demonstrated in this article.
 The theoretically estimated values and
 the reconstructed results shown in the previous section
 are also given.
%
 $\dagger$: estimated for 130 GeV $\mchi$.
}
\label{tab:setup}
\end{center}
\vspace{-0.4cm}
\end{table}

 Secondly,
 all WIMP--nucleon couplings/cross sections
 as well as the ratios between them
 have also been reconstructed with
 only $\sim$ 5\% to $\lsim$ 40\% deviations
 from the input/theoretically estimated values.
 Although the SI WIMP coupling $|f_{\rm p}|$
 estimated with the input (larger) local Dark Matter density
 (lower frame of Figs.~\ref{fig:fp2})
 is {\em underestimated} (Shan \cite{DMDDfp2}),
 one can at least give an upper bound on $|f_{\rm p}|$.
 Meanwhile,
 although the $\armn / \armp$ ratio given in Eq.~(\ref{eqn:ranap_rec})
 is overestimated,
 in Sec.~2.4
 we have demonstrated that
 by combining different methods
 for estimating different (ratios between the) WIMP couplings/cross sections,
 one could in principle observe/confirm the (in)compatibility
 between these results
 and probably correct the reconstructed values.

 Moreover,
 for {\em generating} pseudodata,
 we have used the shifted Maxwellian velocity distribution:
\beq
   f_{1, \sh}(v)
 = \frac{1}{\sqrt{\pi}} \afrac{v}{\ve v_0}
   \bbigg{ e^{-(v - \ve)^2 / v_0^2} - e^{-(v + \ve)^2 / v_0^2} }
\~,
\label{eqn:f1v_sh}
\eeq
 with the Sun's Galactic orbital velocity $v_0 = 230$ km/s;
 $\ve$ is the {\em time--dependent} Earth's velocity
 in the Galactic frame:
\beq
   \ve(t)
 = v_0 \bbrac{1.05 + 0.07 \cos\afrac{2 \pi (t - t_{\rm p})}{1~{\rm yr}}}
\~,
\label{eqn:ve}
\eeq
 the date
 on which the Earth's velocity relative to the WIMP halo
 is maximal
 has been set as $t_{\rm p} = 140$ d.
 Although these values for the astronomical setup
 are {\em non--standard},
 we would like to stress that,
 firstly,
 for using the \amidas\ package and website
 to analyze (real) data sets,
 one needs only the form factors
 for SI and/or SD WIMP--nucleaus cross sections,
 prior knowledge/assumptions about
 the WIMP velocity distribution $f_1(v)$
 and local density $\rho_0$
 (except the estimation of the SI WIMP--nucleon coupling $|f_{\rm p}|^2$)
 are {\em not required}.
 Secondly,
 as shown in the previous section,
 such non--standard values
 would not affect the reconstructed results.

\section{Summary}
 In this article
 I demonstrated the data analysis procedures
 for extrating WIMP properties
 by using theoretically generated pseudodata
 for different target nuclei.
 As an extension as well as the complementarity of
 our earlier theoretical works,
 I combined reconstructed results of
 the (ratios between different) WIMP couplings/cross sections on nucleons
 to estimate each {\em individual} coupling/cross section.
 Hopefully,
 the \amidas\ package and website
 as well as this demonstration
 can help our experimental colleagues
 to analyze their {\em real} direct detection data in the near future
 and to determine (at least rough ranges of) properties of
 halo Dark Matter particles.

\subsection*{Acknowledgments}
 The author appreciates the ILIAS Project and
 the Physikalisches Institut der Universit\"at T\"ubingen
 for kindly providing the opportunity of the collaboration
 and the technical support of the \amidas\ website.
 The author would also like to thank
 the friendly hospitality of the
 National Institute for Nuclear and High Energy Physics (NIKHEF)
 where part of this work was completed.
 This work
 was partially supported
 by the National Science Council of R.O.C.~%
 under contract no.~NSC-99-2811-M-006-031
 as well as by
 the National Center of Theoretical Sciences (South), R.O.C..
%
%

%
\end{document}